# Multi-Band Optical Variability of Blazar 1ES 2344+514 on Diverse Time-Scales

Aykut Özdönmez[1]

[1]Department of Astronomy and Space Science, Faculty of Science, Atatürk University, Erzurum, Türkiye

**Abstract**— This study presents the results of multi-band observations from 2022 to 2024 and Zwicky Transient Facility (ZTF) observations from 2018 to 2023, examining the flux variability of the blazar 1ES 2344+514 on diverse time-scales in the optical bands. The blazar has mild short-term variability (STV) and long-term variability (LTV), with small amplitudes of $\sim$ 0.7 mag and 0.4 mag for the host subtracted- and included-light curves, respectively. The power-enhanced F-test and the nested Analysis of Variance (ANOVA) statistical tests of the six intra-day light curves show that the blazar has no minute-scale variability. The multiband color behavior analysis revealed a moderate redder-when-brighter (RWB) trend on intra-day time scales, while the LTV shows no detectable color behavior. We found a strong correlation between the ZTF optical light curves without any time lag, but no detectable correlations for the optical band emissions. From our periodicity searches using WWZ and LS methods, three significant quasi-periodic oscillation (QPO) signals in the ZTF light curves are found at about 1.02, 1.3, and 2.85 years. The observational results indicate that the blazar 1ES 2344+514 has a complex variability while emphasizing the need for future observations to unravel its underlying mechanisms.

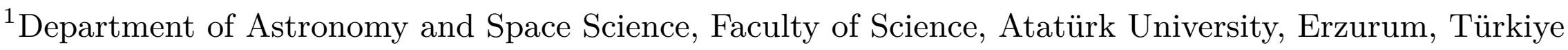

## 1. Introduction

One class of Active Galactic Nuclei (AGNs) called a blazar consists of an accretion disc, a supermassive black hole at the galaxy's centre, and a relativistic jet oriented in the direction of the observer. Blazars show flux variations, strong polarisation, superluminal velocity, non-thermal emission, and high-energy gamma-ray radiation [1,2]. A double-hump structure is seen in the spectral energy distribution (SED) of the blazars. Synchrotron emission from relativistic electrons in the jet's magnetic field produces the low-energy peak, while the inverse Compton process in the MeV–TeV range produces the high-energy peak in the SED [3,4]. Blazars are divided into flat-spectrum radio quasars (FSRQs) and BL Lacertae objects (BL Lacs). Broad emission lines are seen in the optical spectrum of FSRQs, and very faint or nonexistent emission lines are seen in the optical spectrum of BL Lacs [5,6]. Analyzing intra-day variability (IDV), short-term variability (STV), and long-term variability (LTV) of blazar flux is the most common way to study its nature. While STV takes days to months and LTV lasts months to years, Insured Declared Value (IDV) occurs rapidly over minutes to hours (e.g. [7–9]. Many models have been used to try to explain the variations, such as shocks-in-jet [10], gravitational microlensing [11], and variations in Doppler factor [12]. However, there is ongoing debate about many details of the model [13].

[1]aykut.ozdonmez@atauni.edu.tr (Corresponding Author)

Variations in their optical color and spectral characteristics also reveal the emission mechanisms of blazars. The Bluer-When-Brighter (BWB) trend, most commonly observed in BL Lac objects, is often associated with the shock-in-jet hypothesis and variations in the Doppler factor (e.g., [14–17]. Contrarily, additional radiation out from the accretion disk or host galaxy causes the redder-when-brighter (RWB) trend, which is commonly observed in FSRQs (e.g., [18,19]. When the color behavior is not detectable or significant, achromatic behavior is also seen in the blazars [15]. However, it is difficult to fully understand the underlying mechanisms causing these trends due to the complexity of color changes across different time-scales and observational datasets. [20–23]. The interaction between the core engine and the surrounding environment, as well as the jet structure has a significant effect on the color trend and flux variability of blazars.

In many studies, periodic or quasi-periodic flux variability ranging from minutes to decades have been detected in numerous blazars. Many models have been suggested for periodic flux variation-driven mechanisms (e.g. [24–35]). Such models include jet precession, helically moving plasma blobs or shocks within the jet, and binary supermassive black hole systems. Shorter time-scale oscillations, such as those with periodicities of minutes or days, are typically associated with the central emission region, which includes the black hole and accretion disk.

The blazar 1ES 2344+514 is highly variable and an extreme high-frequency-peaked blazar (EHBL), which has a redshift of $z = 0.044$ [36]. Since its discovery in 1995, it has had many GeV-TeV outbursts. Its radio to very high energy emissions have been studied with data from several telescopes, especially during flares [37–42]. There have also been studies of the optical variability of 1ES 2344+514. On September 20, 1996, early observations revealed a 0.08 mag BVR filter microvariability [43]. [44] recorded a 26-minute V-band IDV of 0.14 mag. Many studies have researched LTV and IDVs, but no noteworthy IDV has been found on a time-scale of hours or less [45–48]. However, [49] found a flux variation of $0.69\pm0.16$ mag over 79 min, matching the X-ray variability [50]. LTV was observed in [51], which also included 19 intra-day observations; however, no significant IDV was detected. The authors also reported that the spectral index alpha values of 1.77 and 1.61 remained consistent regardless of whether the host galaxy contribution was included or not. They concluded that the optical emission from this source is likely only affected by non-thermal radiation originating from another region in the relativistic jet. Recently, [48] observed an IDV of $\Delta R = 0.155$ mag over $\Delta T = 12.99$ minutes, indicating a link between magnetic field strength and IDV occurrence. They also reported LTV periodicities of 2.72, 1.61, and 1.05 years.

The scope of this study is to use multi-color observations from intra-day to long-term time-scales to examine the optical flux variations, correlations between various optical bands, color behavior, and periodicity of the blazar 1ES 2344+514 in order to understand its nature and structure.

## 2. Data

Utilising a 60 cm RC robotic telescope (T60) and a 1.0 m RC telescope (T100) at the Scientific and Technological Research Council of Türkiye (TUBITAK) National Observatory (TUG), we carried out optical observations in the BVRI bands between July 2022 and January 2024. We took a total of 1366 frames over the duration of 75 nights of data collection. Depending on the band and brightness of the source, the exposure times ranged from 20 to 180 seconds. We subtracted the bias, applied flat-fielding, and eliminated cosmic rays as part of the usual CCD reduction procedure. As reported in [52], the host galaxy contributes significantly to the total optical flux with 90%. They suggested that the contribution of the host galaxy depends on the aperture radius during photometry, and they determined a host galaxy brightness of 14.90 mag in R-band for a 5” aperture. Thus, we used the same fixed aperture radius for photometry to account for the host contamination. From the chart provided by [53], we chose star C2 as the reference due to its proximity to the blazar’s position, brightness, and color, and stars C1 and 1 were used as comparisons for the instrumental magnitude evaluation. To obtain the flux of the central region, we first corrected the magnitudes of the Galactic extinction using the $A_\lambda$ values obtained from the NASA/IPAC Extragalactic.

Then we converted the R-magnitude to flux density as given in [54]. Thus, we subtracted the contribution of the host galaxy and obtained the source flux density based on the aperture radii and brightness given in [52]. Finally, we determined the flux and magnitude in the R-band, either with the host flux subtracted or included. The R-band flux of the host galaxy is also used to determine the corresponding contribution in the B, V, and I bands for the given galaxy colors [55].

In addition to our observational data, we collected Zwicky Transient Facility (ZTF) light curve data between May 2018 and October 2023 in the gr-bands from the ZTF.

To maintain the ZTF light curve’s quality, we selected 1333 ZTF data points with catflags = 0 and chi< 4. It should be noted that since both the galaxy colors in the ZTF bands and used aperture size are not known, the host subtraction is not applied for ZTF fluxes.

## 3. Flux Variability

Studying the flux variability of blazars at different wavelengths and time-scales, from the shortest to the longest, is crucial to understand the nature of the emission regions and the underlying radiative processes. Our observations of the blazar 1ES 2344+514 allowed us to study the flux and color variability on diverse time-scales, as well as the role of the host galaxy in these phenomena. Figure 1 shows the corresponding intra-day light curves of the multi-band (BVRI) data obtained from all six nights with at least one hour of observations.

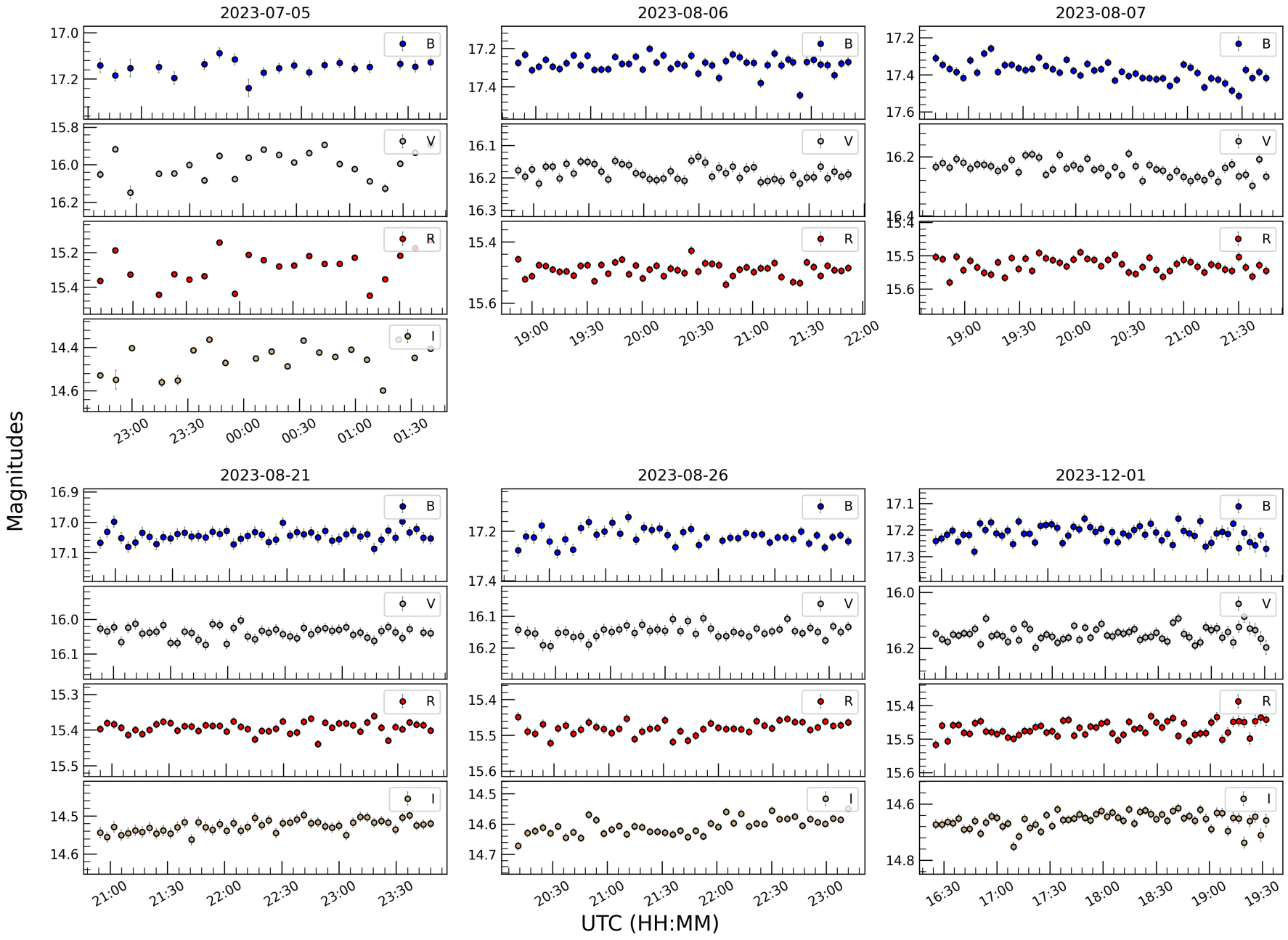

**Figure 1.** Intra-day light curves of blazar 1ES 2344+514 in the optical $BVRI$ bands. For better visualization, the blue, green, gray and dark red circles represent the B-, V-, R- and I-band magnitudes, respectively, with a labeled offset. The dates of the observations are given at the top of each plot

To detect variability in the optical BVRI bands of the blazar 1ES 2344+514, two common statistical tests were used [9, 56–59]: the power-enhanced F-test and the nested analyzes of variance (ANOVA) test. The power-enhanced F-test compares the variance of the blazar's light curve to the total variance of comparison stars using the formula $F_{enh} = \frac{s_{bl}^2}{s_c^2}$ [60]. $s_{bl}^2$ represents the blazar's differential variance, and $s_c^2$ represents the total variance of the comparison stars. The formulas for the degrees of freedom (DOF) are $u_{bl} = N - 1$ and $u_c = k(N - 1)$. For the variability hypothesis to be valid, the $F_{enh}$ is expected to be equal to or greater than the critical value ($F_{critical}$), which is calculated at a 99% confidence level ($\alpha = 0.01$). In our study, we calculated the $F_{\mathrm{enh}}$ values by analyzing the differential light curves from all ten comparison stars and the reference star (field star C2 from the mentioned chart). We then compared these values with the crucial value ($F_c$) for each ID lightcurve. Nested ANOVA, a modified version of ANOVA, is used to assess mean group variations in the blazar's differential light curves by employing multiple reference stars. Using nested ANOVA has the advantage of not relying on a specific comparison star, allowing any available field stars to serve as reference stars for the study. In our investigation, we constructed differential LCs for the blazar using eleven standard stars (including star C2). We created five data points per group from these differential LCs. The mean square of the groups ($MS_G$) and the nested observations within the groups ($MS_{O_G}$) were computed using the methods described in [61] and [56] as well. The resulting ratio, $F = MS_G/MS_{O_G}$, has $a(b-1)$ and $(a-1)$ degrees of freedom in the denominator and numerator, respectively, and is distributed according to a $F$ distribution. Hypothesis tests on the variability of Blazar's light curve were validity tested at a significance level of $\alpha = 0.01$. Table 1 presents the results of the nested ANOVA tests and the $F_{\mathrm{enh}}$-tests. If the $F$-statistic in both tests is equal or greater than the critical value ($F_c$), the ID light curve is considered variable (V); otherwise, it is considered non-variable (NV). The six intra-day LCs showed no significant variability on minute-time scales. Although the standard deviation and mean error of our ID data are both less than 0.03 mag, we could not detect any significant IDV in the LCs similar to the $\Delta R = 0.155$ mag detected by [48], except for a few scatter points. It should be considered that the probability of detecting IDV increases with observational time. [62] found that blazars are variable on intra-day time-scales for $\sim 60 - 65\%$ when observed for $< 6$ hours and $\sim 80 - 85\%$ when observed for more than 8 hours. However, the observational data in our light curve spans one to three hours. Given this discrepancy in observation time, it is possible that any potential variability in the ID LCs of the blazar may not have been fully captured.

**Table 1.** The test results of IDV of the blazar 1ES 2344+51.4

| Obs. date | Band | Average | $t_{\mathrm{obs}}$ | Power-enhanced F-test | | | Nested ANOVA test | | | Status | Variability |
|---|---|---|---|---|---|---|---|---|---|---|---|
| yyyy-mm-dd | | Magnitude | hours | DOF($\nu_1$,$\nu_2$) | $F_{enh}$ | $F_c$ | DOF($\nu_1$,$\nu_2$) | $F$ | $F_c$ | | Amplitude (%) |
| 2023-07-05 | B | 16.38 | 3.10 | (19, 190) | 0.08 | 2.00 | (3, 16) | 0.05 | 5.29 | NV | 6.51 |
| | V | 15.32 | 3.10 | (21, 210) | 0.02 | 1.94 | (3, 16) | 1.68 | 5.29 | NV | 13.30 |
| | R | 14.61 | 3.10 | (21, 210) | 0.06 | 1.94 | (3, 16) | 0.99 | 5.29 | NV | 17.21 |
| | I | 13.83 | 2.96 | (19, 190) | 0.30 | 2.00 | (3, 16) | 0.91 | 5.29 | NV | 12.89 |
| 2023-08-06 | B | 16.45 | 3.00 | (48, 480) | 0.56 | 1.58 | (8, 36) | 0.48 | 3.05 | NV | 10.62 |
| | V | 15.42 | 3.00 | (47, 470) | 0.69 | 1.59 | (8, 36) | 2.70 | 3.05 | NV | 3.66 |
| | R | 14.72 | 3.00 | (47, 470) | 0.93 | 1.59 | (8, 36) | 2.56 | 3.05 | NV | 5.23 |
| 2023-08-07 | B | 16.49 | 3.02 | (48, 480) | 1.24 | 1.58 | (8, 36) | 5.33 | 3.05 | NV | 10.88 |
| | V | 15.45 | 3.02 | (48, 480) | 0.55 | 1.58 | (8, 36) | 3.49 | 3.05 | NV | 4.81 |
| | R | 14.74 | 3.02 | (48, 480) | 0.63 | 1.58 | (8, 36) | 1.31 | 3.05 | NV | 4.06 |
| 2023-08-21 | B | 16.33 | 2.89 | (47, 470) | 0.57 | 1.59 | (8, 36) | 0.71 | 3.05 | NV | 3.98 |
| | V | 15.35 | 2.89 | (46, 460) | 1.02 | 1.59 | (8, 36) | 1.71 | 3.05 | NV | 3.23 |
| | R | 14.67 | 2.89 | (47, 470) | 1.12 | 1.59 | (8, 36) | 0.97 | 3.05 | NV | 3.76 |
| | I | 13.87 | 2.89 | (47, 470) | 0.39 | 1.59 | (8, 36) | 4.11 | 3.05 | NV | 3.07 |
| 2023-08-26 | B | 16.42 | 2.93 | (41, 410) | 0.80 | 1.63 | (7, 32) | 0.98 | 3.26 | NV | 6.29 |
| | V | 15.40 | 3.02 | (43, 430) | 0.74 | 1.62 | (7, 32) | 2.05 | 3.26 | NV | 3.89 |
| | R | 14.71 | 3.02 | (43, 430) | 0.85 | 1.62 | (7, 32) | 2.72 | 3.26 | NV | 3.29 |
| | I | 13.91 | 3.02 | (43, 430) | 0.51 | 1.62 | (7, 32) | 30.67 | 3.26 | NV | 6.26 |
| 2023-12-01 | B | 16.41 | 3.10 | (60, 600) | 0.89 | 1.51 | (11, 48) | 1.49 | 2.64 | NV | 5.32 |
| | V | 15.40 | 3.11 | (60, 600) | 1.06 | 1.51 | (11, 48) | 0.90 | 2.64 | NV | 5.18 |
| | R | 14.71 | 3.11 | (60, 600) | 0.96 | 1.51 | (11, 48) | 1.32 | 2.64 | NV | 3.86 |
| | I | 13.94 | 3.11 | (60, 600) | 1.02 | 1.51 | (11, 48) | 1.79 | 2.64 | NV | 6.63 |

Figure 2 displays the long-term optical light curves for the ZTF gr- and BVRI-bands, based on all observational data. The analyzes for BVRI data was performed with and without host subtraction, while the analyzes for ZTF data was done only for fluxes with host galaxy light. Table 2 provides a summary of the LTV light curves, including the minimum, maximum, and mean magnitudes. In addition, we estimated the variability amplitude ($A$) as given by [63], in which it is defined to quantify the actual variation in a light curve after correcting for the observational errors. The formula for the variability amplitude ($A$) is $A = 100 \times \sqrt{(A_{\max} - A_{\min})^2 - 2\sigma^2}$ %. Here, $A_{\max}$ represents the maximum magnitude, $A_{\min}$ represents the minimum magnitude of the LTV light curve, and $\sigma$ denotes the average error.

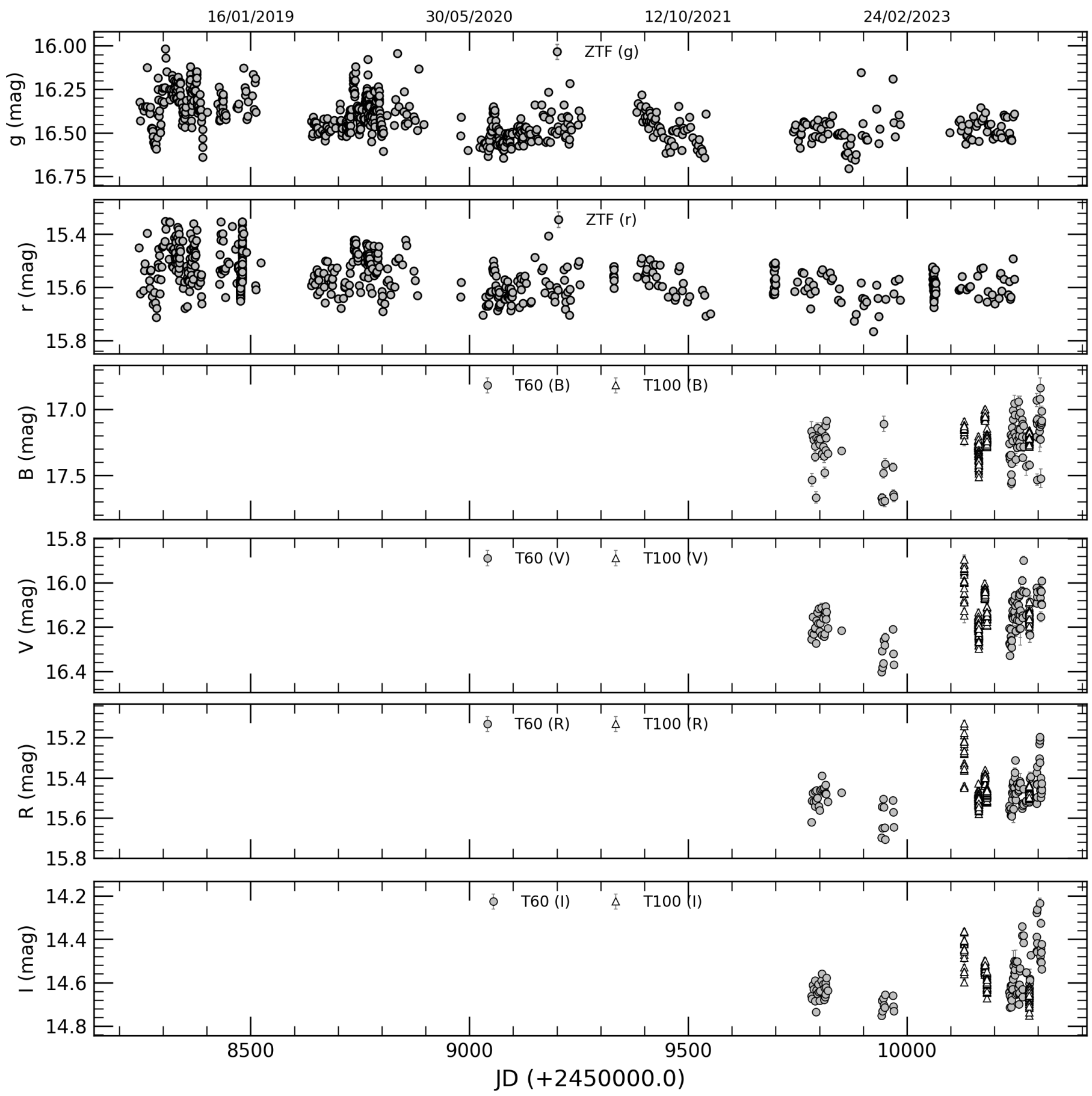

**Figure 2.** Long-term light curves of the blazar 1ES 2344+51.4 in the optical ZTF gri bands and $BVRI$ bands. The BVRI light curves are the host-subtracted but ZTF light curves includes host contribution

**Table 2.** Results of the LTV analyzes of the blazar. Lower given values for BVRI bands includes host galaxy light

| Band | Brightest magnitude/MJD | Faintest magnitude/MJD | Average magnitude | Variability amplitude (%) |
|---|---|---|---|---|
| B | $16.836 \pm 0.073$ / 60305.24014 | $17.699 \pm 0.033$ / 59944.20939 | $17.232 \pm 0.026$ | 86.252 |
| | $16.216 \pm 0.073$ | $16.619 \pm 0.033$ | $16.420 \pm 0.026$ | 40.108 |
| V | $15.894 \pm 0.014$ / 60131.53178 | $16.403 \pm 0.015$ / 59942.21587 | $16.147 \pm 0.016$ | 50.843 |
| | $15.266 \pm 0.014$ | $15.522 \pm 0.015$ | $15.400 \pm 0.016$ | 25.477 |
| R | $15.128 \pm 0.016$ / 60131.57453 | $15.707 \pm 0.013$ / 59950.19463 | $15.463 \pm 0.012$ | 57.815 |
| | $14.525 \pm 0.016$ | $14.818 \pm 0.013$ | $14.703 \pm 0.012$ | 29.305 |
| I | $14.235 \pm 0.024$ / 60303.27906 | $14.752 \pm 0.015$ / 60280.21487 | $14.585 \pm 0.014$ | 51.582 |
| | $13.698 \pm 0.024$ | $13.983 \pm 0.015$ | $13.897 \pm 0.014$ | 28.414 |
| zg | $16.018 \pm 0.014$ / 58304.9748 | $16.703 \pm 0.016$ / 59866.72145 | $16.415 \pm 0.015$ | 68.564 |
| zr | $15.351 \pm 0.011$ / 58304.90155 | $15.767 \pm 0.011$ / 59922.78525 | $15.555 \pm 0.011$ | 41.608 |

The long-term light curve exhibits varying behavior, with amplitudes ranging from 25% to 70% depending on the band. Host subtraction increases the variability amplitude by about a factor of two. The variability in brightness can vary by up to 0.4 mag within a few days, or even within a single day, for light curves. During our observations, the brightest magnitude was measured to be $R = 14.525$ mag, while the faintest brightness was $R = 14.818$ mag. Meanwhile, the mean R-band magnitude was $14.703 \pm 0.012$ mag, with a calculated variability amplitude of 29%. The ZTF observations show the brightest magnitude at $r = 15.351$ and the faintest at $r = 15.767$ mag, resulting in a mean magnitude of 16.415 and a variability amplitude of 42%. Throughout both observations, no flare was detected.

### 3.1. The Multiband Color Behavior and the Correlation Analyzes

Blazars show rapid variations in both flux and spectral characteristics, which are caused by the combination of thermal emission from the accretion disc and non-thermal synchrotron radiation from the relativistic jet. Given that the color variability of blazars mirrors their spectral variations, we conducted a comprehensive analyzes of the correlations between magnitudes and colors through intra-day and long-term time-scales. For the intra-day time-scales, we focused on the $B - R$ and $V - R$ indices with respect to the R magnitude. The color-magnitude plots are shown in Figure 3, and the results of the linear regressions are listed in Table 3. We identified moderate correlations ($r < -0.6$ and $p < 0.05$) for both the $B - R$ and $V - R$ color indices, regardless of the presence of host contributions. The negative slope indicates RWB trend.

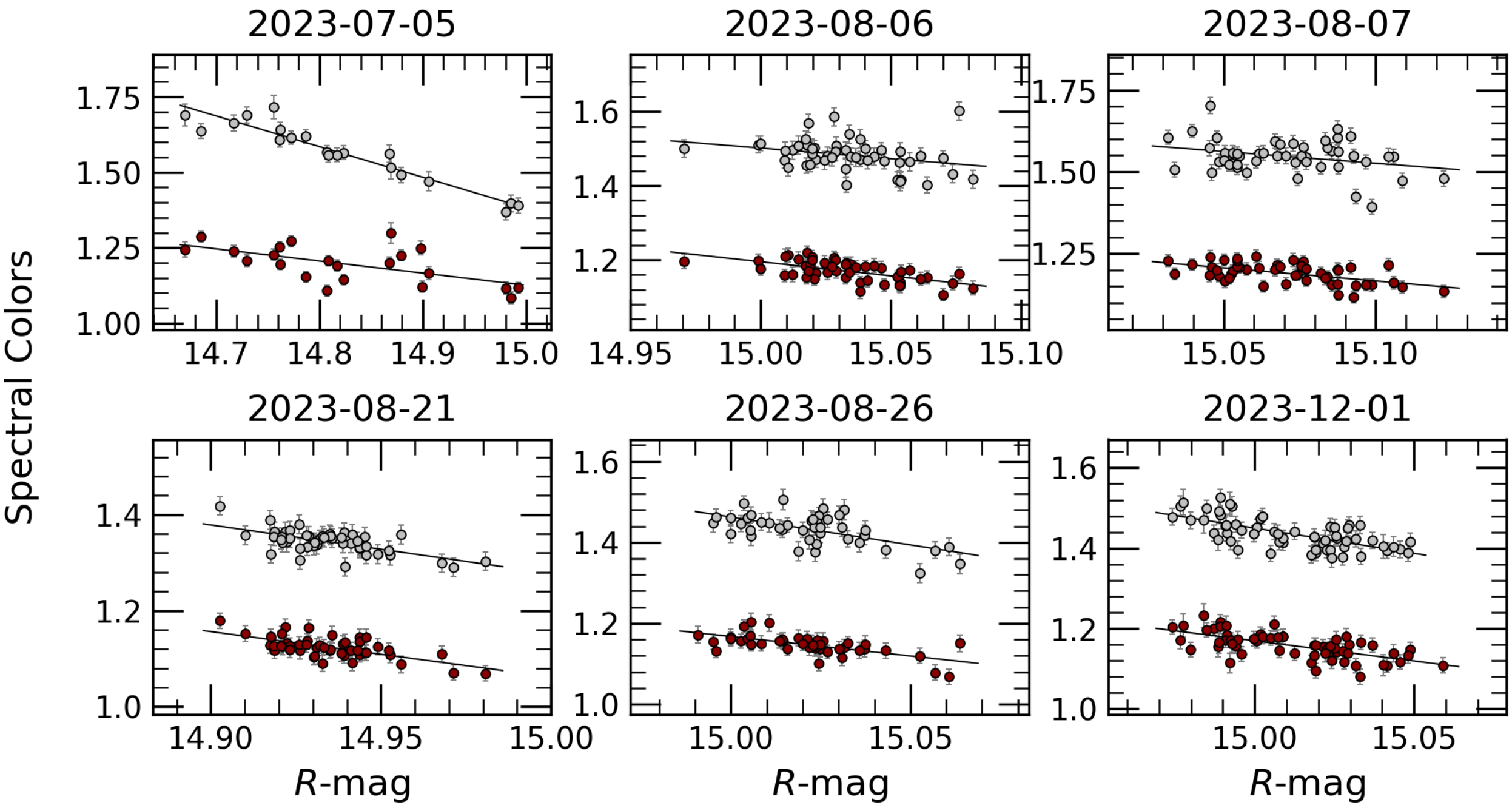

**Figure 3.** The intra-day color–magnitude plots of the blazar 1ES 2344+51.4 obtained from host-subracted fluxes. The date are presented above each plot. The silver and dark red dots represent $B - R$ and $V - R$ color indices. Black lines represent linear fits. For better representation, the offset value of 0.6 are added to $V - R$ colors

**Table 3.** Correlation between the color indices and R-band magnitude for multiband IDV data

| Date | color | slope | r | p |
|---|---|---|---|---|
| 2023-07-05 | $B-R$ | $-1.02 \pm 0.08$ | -0.95 | 1.22e-10 |
| | $V-R$ | $-0.40 \pm 0.12$ | -0.60 | 2.85e-03 |
| | $^*B-R$ | $-1.01 \pm 0.07$ | -0.96 | 3.08e-11 |
| | $^*V-R$ | $-0.40 \pm 0.12$ | -0.61 | 2.69e-03 |
| 2023-08-06 | $B-R$ | $-0.57 \pm 0.27$ | -0.30 | 4.16e-02 |
| | $V-R$ | $-0.76 \pm 0.14$ | -0.62 | 3.17e-06 |
| | $^*B-R$ | $-0.60 \pm 0.25$ | -0.33 | 2.13e-02 |
| | $^*V-R$ | $-0.76 \pm 0.15$ | -0.61 | 4.17e-06 |
| 2023-08-07 | $B-R$ | $-0.72 \pm 0.34$ | -0.30 | 3.94e-02 |
| | $V-R$ | $-0.80 \pm 0.18$ | -0.55 | 4.92e-05 |
| | $^*B-R$ | $-0.75 \pm 0.31$ | -0.33 | 2.03e-02 |
| | $^*V-R$ | $-0.80 \pm 0.18$ | -0.55 | 5.01e-05 |
| 2023-08-21 | $B-R$ | $-1.01 \pm 0.18$ | -0.64 | 1.10e-06 |
| | $V-R$ | $-0.94 \pm 0.16$ | -0.66 | 4.96e-07 |
| | $^*B-R$ | $-1.01 \pm 0.18$ | -0.63 | 1.42e-06 |
| | $^*V-R$ | $-0.94 \pm 0.17$ | -0.65 | 9.31e-07 |
| 2023-08-26 | $B-R$ | $-1.37 \pm 0.29$ | -0.61 | 2.09e-05 |
| | $V-R$ | $-0.96 \pm 0.17$ | -0.65 | 1.94e-06 |
| | $^*B-R$ | $-1.37 \pm 0.28$ | -0.62 | 1.46e-05 |
| | $^*V-R$ | $-0.96 \pm 0.18$ | -0.64 | 2.82e-06 |
| 2023-12-01 | $B-R$ | $-1.25 \pm 0.18$ | -0.67 | 5.42e-09 |
| | $V-R$ | $-1.00 \pm 0.15$ | -0.65 | 1.43e-08 |
| | $^*B-R$ | $-1.24 \pm 0.18$ | -0.68 | 2.57e-09 |
| | $^*V-R$ | $-1.00 \pm 0.15$ | -0.64 | 2.24e-08 |

* The results are obtained from for the colors including host contribution

For the long-term multiband data from ZTF and our observations, the color magnitude plots are shown in Figure 4. We observed very weak correlations for all color indices, with a correlation coefficient of $|r| < 0.3$, and the slope errors are similar to the slope values (Table 4). Thus, the color relation for LTV is not signficant; i.e. achromatic trend.

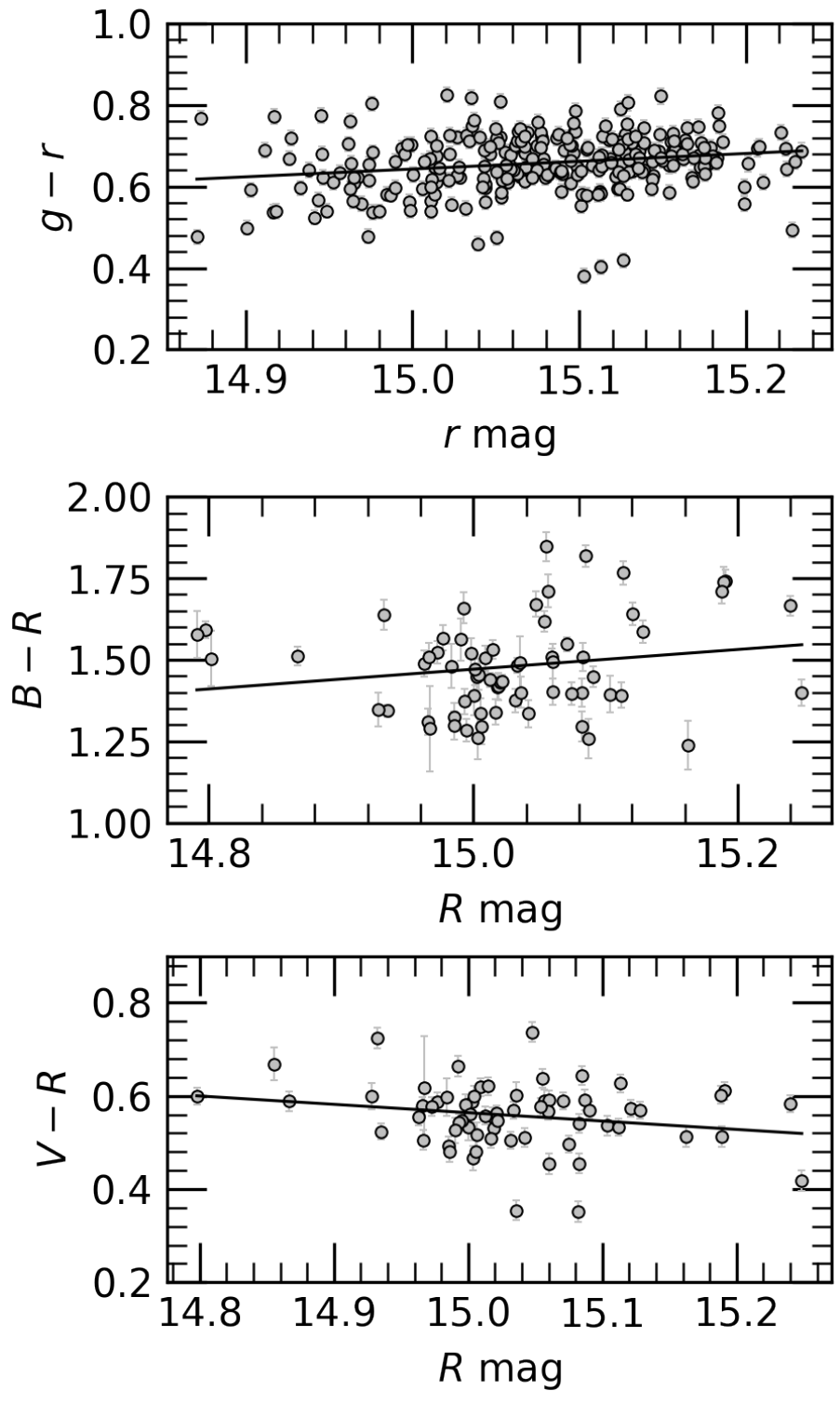

**Figure 4.** The color-magnitude diagram of the blazar 1ES 2344+51.4 on long-term time-scale obtained from ZTF gr-band magnitudes which includes host light and from host subtracted BVR magnitudes Black lines represent the linear fits

**Table 4.** The results of the linear regression between the colors and the corresponding magnitude on the long-term time-scales

| Model | slope | intercept | r-value | p-value |
|---|---|---|---|---|
| $R$ vs $B-R$ | $+0.303 \pm +0.201$ | $-3.070 \pm 3.024$ | +0.183 | 1.373e-01 |
| $R$ vs $V-R$ | $-0.179 \pm +0.104$ | $+3.255 \pm 1.563$ | -0.211 | 8.927e-02 |
| $^{*}r$ vs $g-r$ | $0.191 \pm 0.055$ | $-2.221 \pm 0.834$ | 0.206 | 6.43e-04 |
| $^{*}R$ vs $B-R$ | $+0.163 \pm 0.184$ | $-0.916 \pm 2.623$ | +0.110 | 3.774e-01 |
| $^{*}R$ vs $V-R$ | $-0.176 \pm 0.106$ | $+3.086 \pm 1.510$ | -0.204 | 1.011e-01 |

* The results are obtained from for the fluxes including host contribution

Understanding the correlations between different spectral bands is critical because any identified time lag implies spatial differences between the corresponding emission regions. Thus, we used the Discrete Correlation Function (DCF), which is an effective statistical tool for investigating unevenly distributed time series of multiband light curves ( [64] and references therein). The calculation of the DCF involves using the unbinned DCF (UDCF), which is expressed as:

$$UDCF_{ij}(\tau) = \frac{(a_i - \bar{a})(b_j - \bar{b})}{\sqrt{(\sigma_{a^2} - e_{a^2})(\sigma_{b^2} - e_{b^2})}} \tag{3.1}$$

Here, $\bar{a}$ and $\bar{b}$ are the mean values of the respective time-series datasets, while $\sigma_{a,b}$ and $e_{a,b}$ are their standard deviations and errors. The time delay between two data points is denoted by $\Delta t_{ij} = (t_{bj} - t_{ai})$. The DCF is then found by taking the average of the UDCF values over the interval $\tau - \frac{\Delta\tau}{2} \leq \tau_{ij} \leq \tau + \frac{\Delta\tau}{2}$ as given in [65]:

$$DCF(\tau) = \frac{\sum_{k=1}^{m} \mathrm{UDCF}_k}{M} \tag{3.2}$$

Here, "M" denotes the number of pairwise time lag values within the specified $\tau$ interval.

In the examination of the long-term light curves, weighted mean magnitudes and mean Julian dates (JD) were computed from nightly binned observations. The DCF analyzes (see Figure 5) were applied to each pair combination of the nightly binned ZTF $gri$ and optical $VRI$ long-term light curves using a time binning value of 10 days. We found a strong correlation between the g- and r-band light curves (LCs), but this correlation is not significant for the other multi-band pairs. For all pairs, the maximum DCF values were found to be at $\tau = 0$ days, which implies that the optical emission regions are likely co-spatial.

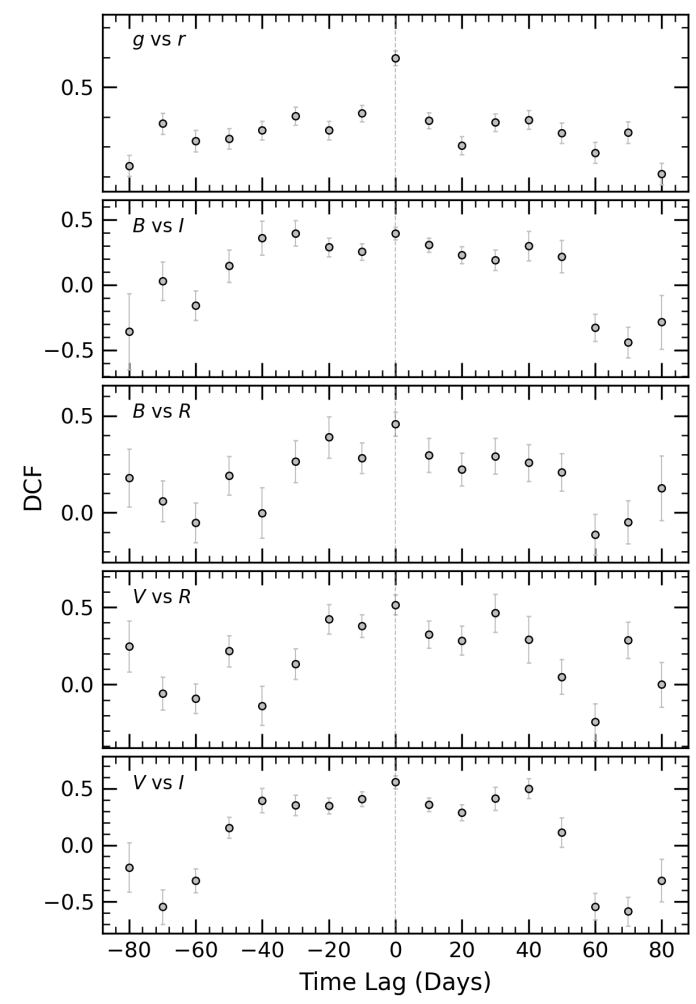

**Figure 5.** Cross-correlation analyzes for the ZTF $gri$ and optical $BVRI$ bands using Discrete Correlation Function for the entire monitoring period

### 3.1.1. Optical Quasi-Periodic Ossilations

In order to identify any underlying periodicity or optical quasi-periodic ossilation (QPO) of flux variability on long-term time-scales, we used the ZTF light curves of 1ES 2344+514, which consist of longer and continuous observational data. We employed two widely recognised methods: the Lomb-Scargle (LS) periodogram [66, 67], and the Weighted Wavelet Z-transform (WWZ; [68]). The LS periodogram is a Fourier transform method used to detect periodic patterns in time series data by integrating sinusoidal components based on likelihood minimisation. Utilising the Python Astropy package, the LS periodogram provided a detailed analyzes of quasi-periodic signals within the corresponding time series. The WWZ method represents an enhanced version of the wavelet transform, designed to detect periodic or quasi-periodic signals in unevenly sampled time series data and evaluate their consistency over the observation period. The signal is simultaneously decomposed into both the frequency and time domains, and this method employs wavelet functions to model the observation rather than sinusoidal components. We performed a thorough WWZ analyzes of the light curves over the entire observation period. Accurately estimating the significance of potential detections is crucial to avoid misinterpreting noise-induced peaks as periodic signals in the power spectrum. To ensure robust analyzes, we generated 10,000 simulated light curves reflecting the power spectral density of the blazar light curve, following the methodology outlined in [69]. By performing identical period analyzes using WWZ and LS on the simulated light curves, as was done on the Blazar light curve, we obtained significance levels of 99% and 99.9% of red noise for each frequency.

The resulting plots of the WWZ and LS analyzes for the ZTF $r$-band light curve are shown in Figure 6, including the WWZ power map corresponding to the periodicity and time, the periodogram of the time-averaged WWZ power, and the LS periodogram. The WWZ analyzes revealed two significant signals at a periodicity of $373^{+89}_{-74}$ and $1041^{+351}_{-127}$ days in the WWZ periodogram, spanning the entire observation time in the WWZ power map and exceeding the significant level at 99%, almost at 99.9%. The LS periodogram also exhibited two similar peak signals at $479^{+173}_{-74}$ days and $1065^{+258}_{-193}$ days. First peak surpasses the significance level at 99%, but significance of the second peak is at about 98% . Notably, similar results were obtained for the ZTF g-band light curve. We also checked the periodicity for our BVRI light curves only, but found no significant signal for the STV periodicity.

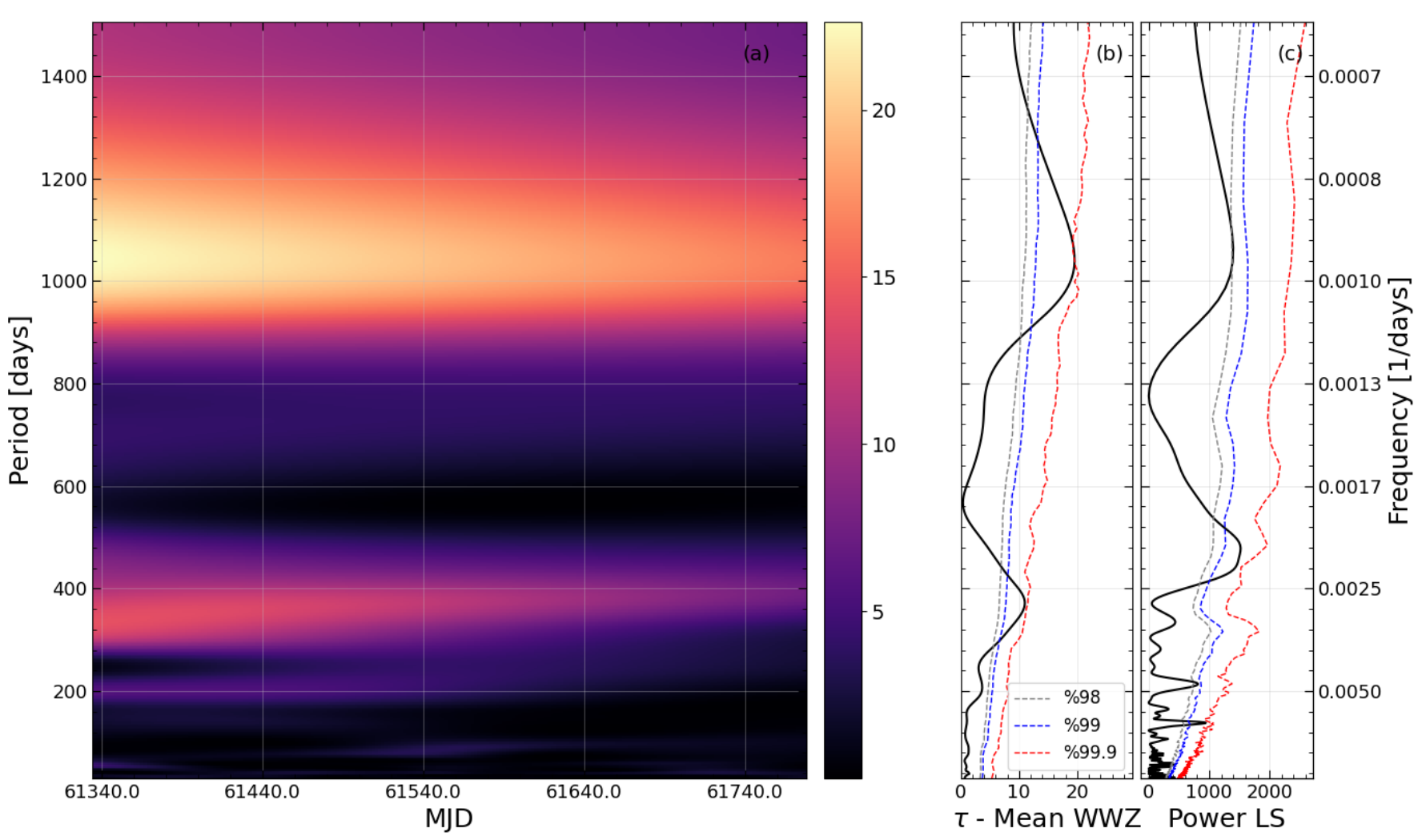

**Figure 6.** (a) 2D plane contour plot of the WWZ power over time and frequency domain, (b) Time averaged WWZ power periodogram, and (c) the Lomb-Scargle periodogram of the combined ZTF $r$ band light curve. Dotted brown and red lines represents the significance level of 98, 99 and 99.9 percent, respectively

## 4. Conclusion

This study presents the results of our multi-band observations from 2022 to 2024 and ZTF observations from 2018 to 2023, examining the flux variability of the blazar 1ES 2344+514 on diverse time-scales in optical bands. The blazar 1ES 2344+514 has mild STV and LTV, with a small amplitude of $\sim 0.7$ mag and 0.4 mag for host subtracted- and included-light curves, respectively. During whole observation period, no flare was recorded. For a 10-day time binning with the optical datasets in this study, the obtained correlation between ZTF LCs suggests that the optical emission region is likely to be co-spatial or that any spatial differences are too small to detect. Statistical analyzes on the six multi-band ID light curves show that the blazar has no minute-scale variability. Moreover, the IDV has been reported only for a few nights in the literature, as well. The detected IDV fraction for this blazar is significantly less than that of typical BL Lac objects [70–72]. As a point source, a substantial emission from the host galaxy is known to dilute the underlying IDV [48]. Even though the host galaxy contribution is subtracted from the total flux, we didn't detect IDV. The absence of IDV in a blazar can be attributed to the presence of a powerful magnetic field that prevents or delays the occurrence of instability. This instability is likely responsible for the significant change in the jet's morphology, which could explain the observed rapid variability [73]. This explanation is plausible for Blazar 1ES 2344+514 because to its magnetic field strength beyond the crucial value, as reported by [48]. It is worth noting that 1ES 2344 + 514 is classified as a HBL source, but low-energy-peaked blazars are more likely to exhibit IDV than HBLs [70, 72].

The multiband color behavior analyzes indicated that a typically moderate and detectable RWB trend for intra-day time-scales exists, but no detectable color behavior is found for the LTV. The RWB trend is more commonly detected in FSRQs and is thought to be caused by additional contributions from the accretion disc or other emission regions (e.g. [18]). The color trend of the blazars is determined by several parameters, including the synchrotron peak frequency, the width of optical wavelengths, and the intensity of thermal blue emission compared to jet emission. We know that the host galaxy mostly contaminates thermal emissions [16]. Therefore, an RWB trend and color saturation can be observed in the blazar when the jet and other emission regions, such as the host galaxy and the accretion disc, contribute varying amounts of thermal emission over time. Variations in the Doppler factor are commonly used to explain achromatic behavior. The geometric scenarios provide the most plausible explanation for these differences.

Blazars frequently exhibit periodic variation in their light curves, such in case of 1ES 2344+514. Our results suggest three significant QPO signals in ZTF light curves at around 1.02, 1.3, and 2.85 years. These signals are very similar with the reported periodicities of 1.05, 1.61, and 2.72 years by [48], in which binary black hole system is considered for the explanation of these periodicities. In case of blazars, it is proposed that the most likely reasons for the year-scale periodicity are binary black hole in the central region, and helical-jet structure. The presence of closely orbiting binary black holes or warped accretion discs can cause the jet precession phenomenon when a blazar is involved in a binary SMBH system (e.g. [74–76]. Depending on the angle of the jet axis to the observer's line of sight and the jet's Lorentz factor, this mechanism is likely to produce QPOs at year-like time-scales, as found in this study. On the other hand, helical structures are common in blazars [77, 78]. The Doppler boosting effect due to helical or non-ballistic motions of relativistic blobs or shocks within blazar jets could be responsible for the QPO in blazar LCs (e.g. [79, 80]. The simplest leptonic one-zone model suggests that the viewing angle of the blob with respect to the line of sight ($\phi_{obs}$) changes periodically with time due to the postulated helical motion of the blob, and the Doppler factor ($\delta$) varies with viewing angle as $\delta = 1/[\Gamma(1 - \beta \cos\theta(t))]$. Here, $\Gamma = 1/\sqrt{1 - \beta^2}$ is the bulk Lorentz factor of the blob motion with $\beta = \nu_{jet}/c$. Given this scenario, the periodicity in the rest frame of the blob is given by $P_{rf} = P_{obs}/(1 - \beta \cos\psi \cos\phi)$. For typical values of the pitch angle of the helical path $\phi = 2°$, the angle of the jet axis with respect to the line of sight $\psi = 2°$, and Doppler factor $\Gamma = 10$, the periodicities in the rest frame of the blob are calculated as $\sim$ 164, 209, and 458 years for the observed periodicities (see for methodology [81–83]). During these rest frame periodicities, the blob traverses distances as $D = c\beta P_{rf} \cos\phi \simeq$ 50, 64, and 139 pc, respectively. It is physically more plausible that the jet has a significantly lower curvature, which explains these length scales and is consistent with the observed periodogram (see for discussion [84]).

To improve our understanding of the complex variability of the blazar 1ES 2344+514, simultaneous multi-wavelength observations as well as the determination of host contamination over time, complemented by theoretical modelling on the underlying mechanisms of blazar variability, are needed.

## Author Contributions

The author read and approved the final version of the paper.

## Conflicts of Interest

The author declares no conflict of interest.

## Ethical Review and Approval

No approval from the Board of Ethics is required.

## Acknowledgement

This study was supported by Scientific and Technological Research Council of Turkey (TUBITAK) under the Grant Number 121F427. The authors thank to TUBITAK for their supports. We thank the team of TUBITAK National Observatory (TUG) for a partial support in using the T60 and T100 telescopes with project numbers: 22AT60-1907 and 23AT100-2006.